\documentclass[
	amsmath,
	amssymb,
	pra,
	reprint,
	superscriptaddress
	]{revtex4-1}

\usepackage{graphicx}
\usepackage{dcolumn}
\usepackage{bm}
\usepackage{hyperref}
\hypersetup{					
	plainpages=false,			
	colorlinks=false,			
	pdfborder={0 0 0},			
	breaklinks=true,			
	bookmarksnumbered=true,		%
	bookmarksopen=true			%
}
\usepackage{physics}
\usepackage{color} 
\usepackage{mathrsfs} 
\usepackage{siunitx}

\definecolor{color_todo}{RGB}{214,45,32}

\definecolor{color_quest}{RGB}{0,87,231}

\definecolor{color_change}{RGB}{255,167,0}

\newcommand{\ie}{i.\,e.\ }
\newcommand{\eg}{e.\,g.\ }
\newcommand{\expp}[1]{\operatorname{e}^{#1}}
\newcommand{\trans}{\text{T}}
\newcommand{\Svec}{\vb{S}}
\newcommand{\Xvec}{\vb{X}}
\newcommand{\Sigmavec}{\Sigma}
\newcommand{\qfi}[2]{\operatorname{I}\qty(#1;#2)}
\newcommand{\QFI}{\operatorname{I}}
\newcommand{\Nth}{{N_\text{th}}}
\newcommand{\nbar}{{\bar{n}}}
\newcommand{\eq}{Eq.\,}
\newcommand{\estimator}{\hat{\theta}_\text{est}}
\newcommand{\ketf}[2]{\ket{#1}_{\!#2}}
\newcommand{\braf}[2]{{\hspace{-14pt}\phantom{\ket{j}}}_{#2}\!\!\bra{#1}}
\newcommand{\braketf}[4]{{\hspace{-14pt}\phantom{\ket{j}}}_{#3}\!\!\braket{#1}{#2}_{\!#4}}
\newcommand{\R}[4]{R_{#1 #2} \qty(#3,#4)}
\newcommand{\ch}{\mathcal{C}_r}
\newcommand{\sh}{\mathcal{S}_r}
\newcommand{\sApp}{see\ Appendix.\ }

\allowdisplaybreaks
\begin{document}

\title{Quantum parameter-estimation of frequency and damping of a harmonic-oscillator}

\author{Patrick Binder}
\affiliation{Institute for Theoretical Physics, Tübingen University , 72076 Tübingen, Germany}
\affiliation{BioQuant Center, Im Neuenheimer Feld 267, 69120 Heidelberg, Germany} 
\affiliation{Institute for Theoretical Physics, Heidelberg University, Philosophenweg 19, 69120 Heidelberg, Germany} 
\author{Daniel Braun} 
\email{Author to whom correspondence should be addressed: daniel.braun@uni-tuebingen.de}

\affiliation{Institute for Theoretical Physics, Tübingen University , 72076 Tübingen, Germany}

\date{\today}

\begin{abstract}	
	We determine the quantum Cramér-Rao bound for the precision
        with which the oscillator frequency and damping constant of a
        damped quantum harmonic oscillator in an arbitrary Gaussian state can be estimated. This goes
        beyond standard quantum parameter estimation of a single mode
        Gaussian state for which typically a mode of fixed frequency
        is assumed. We present a scheme through which the frequency
        estimation can nevertheless be based on the known results for
        single-mode quantum parameter estimation with Gaussian states.
        Based on these results, we investigate {the optimal
          measurement time}. For measuring the oscillator frequency,
        our results unify previously known partial results and
        constitute an   
        explicit solution for a general single-mode Gaussian state.         
          Furthermore, we show that with existing carbon nanotube resonators (see J. Chaste et al.~Nature Nanotechnology 7, 301 (2012))   
it should be possible to achieve a {mass sensitivity 
          of the order of an electron mass $\si{\hertz^{-1/2}}$.}
\end{abstract}


\maketitle

\section{\label{sec:Introduction}Introduction}
The harmonic oscillator is one of the most important model systems in
all of physics.  It is exactly solvable, both classically and quantum
mechanically, and plays a fundamental role in quantum field theories,
where its elementary excitations can be identified with e.g.~photons or
phonons. The harmonic oscillator arises
as low-amplitude limit of a much wider class of non-harmonic
oscillators, and its regular motion is at the basis of time- and
frequency measurements{.}
Indeed, the most precise  measurements of a physical quantity
are often achieved when transducing their variations into frequency
changes.  It is therefore of utmost importance to figure out how
precisely the two {characteristic} quantities of a harmonic
oscillator, namely its frequency and its damping can be measured in
principle. A partial answer was provided in \cite{braun2011ultimate},
where the quantum Cram\'er-Rao bound (QCRB) for the frequency measurement of
an undamped harmonic oscillator in an arbitrary pure quantum state was
calculated.  The QCRB is the ultimate lower bound for the uncertainty
with which a parameter can be estimated.  It is optimized over all
possible (POVM-)measurements (POVM={positive} operator-valued measure, a
class of measurements that includes but is more general than the usual
projective von Neumann measurements), and over all data-analysis
procedures (in the sense of unbiased estimator functions of the
measurement results alone). It becomes relevant when all technical
noise sources are eliminated, and only the noise inherent in the
quantum state remains.  Importantly, the QCRB can be saturated in the
limit of a large number of measurements.

A damped harmonic oscillator leads, however, naturally to
mixed quantum states, and for those the calculation of the QCRB is
much more difficult than for pure states, owing to the need to diagonalize the density 
operator in an infinitely dimensional Hilbert space.  In \cite{PhysRevA.93.013848} an attempt was made to obtain the QCRB for the frequency of a kicked
and damped oscillator \cite{PhysRevA.89.023849}, by using the formulas for Gaussian states.
Indeed, in \cite{PhysRevA.88.040102} the exact QCRB was found for any
of the five parameters that uniquely fix an 
{arbitrary} Gaussian state of a harmonic oscillator. However, those
formulas were derived for an oscillator of fixed frequency, and they
cannot be directly applied for frequency estimation.  Doing so would
amount to considering the {Hamiltonian} $H=\hbar\omega a^\dagger a$ as
a generator of a phase shift, i.e.~the unknown parameter $\omega$
multiplies a hermitian generator, whose variance gives, up to a factor
4, the pure state quantum Fisher information (QFI). However, this
ignores that the annihilation- and creation operators depend
themselves on $\omega$.  That they do so is most easily seen by writing them in
the Fock-basis and realizing that the wave-functions corresponding to
the energy eigenstates depend on $\omega$ through the oscillator
length. Physically, ignoring the $\omega-$dependence of $a,a^\dagger$
hence implies that one neglects the $\omega-$dependence of the
energy-eigenstates, which is particularly important at small times, i.e.~much smaller than the period of the oscillator. 
\\
One might then think that calculating the QCRB for the damped harmonic
oscillator is a hopeless endeavor if the formulas for the Gaussian
states cannot be applied, and the state is not already diagonalized.
Here we show, however, that there is a 
well-defined procedure that allows one to use those formulas
nevertheless for the large and experimentally most relevant class of
initial Gaussian states, by carefully incorporating the consequences
of a change of frequency.  This allows us to fully solve the problem
of parameter estimation of a (weakly) 
damped harmonic oscillator,
described by a Lindblad-master equation.

\section{General framework\label{sec:II}}
We start by briefly describing the dynamics of a damped harmonic
oscillator. Afterwards we review the  closed-form expression for the
general quantum Fisher information (QFI) for single-mode Gaussian
states \cite{PhysRevA.88.040102}. 
\subsection{Dynamics} 
We consider a quantum harmonic oscillator with bare frequency
$\omega$ weakly coupled to a Markovian environment. Assuming the
validity of the Born-Markov approximation and the rotating-wave
approximation, the density
matrix $\rho$ of the oscillator evolves   
according to the master equation (ME) \cite{PhysRevA.4.739,PhysRevA.30.1525}
\begin{eqnarray}\label{eq:ME}
	\dv{\rho}{t} 
		&=&-i\omega\big[\hat{a}^\dagger\hat{a},\rho\big]+\frac{\gamma}{2}\nbar \qty(2 \hat{a}^\dagger \rho\hat{a} - \hat{a}\hat{a}^\dagger \rho - \rho\hat{a}\hat{a}^\dagger)\nonumber\\
		& &+ \frac{\gamma}{2}\qty(\nbar+1)\qty(2 \hat{a}\rho\hat{a}^\dagger -\hat{a}^\dagger\hat{a}\rho-\rho\hat{a}^\dagger\hat{a}),
\end{eqnarray}
where we introduced the mean thermal photon number of the bath $\nbar
= \qty(\expp{x}-1)^{-1}$ 
at frequency $\omega$, dimensionless inverse temperature $x\equiv
\hbar\omega/k_\textsc{B} T$, the damping constant $\gamma$.

By introducing the quadrature operator $\Xvec = \qty(\hat{q},\hat{p})^\trans$, the three-dimensional vector $\Svec(t) = (M\omega\sigma_{qq}, \sigma_{pp}/M\omega, \sigma_{pq})^\trans$, where $\sigma_{AB} \equiv 1/2\times\ev{A B + B A}-\ev{A} \ev{B}$ and by using the ME \eqref{eq:ME} one finds equations of motion \cite{Isar2006}:
\begin{subequations}
	\begin{eqnarray}
	\dv{\ev{\Xvec}\!(t)}{t} 
		&=& G \ev{\Xvec}\!(t),\label{eq:Chap2-1.ord-eq-of-motion}\\
	\dv{\Svec (t)}{t} 
		&=& K \Svec (t)+ 	\Svec_{\text{inh}},\label{eq:Chap2-2.ord-eq-of-motion}
	\end{eqnarray}
where
	\begin{equation}
	G 
		= \begin{pmatrix} -\gamma/2 & 1/M\\ -M\omega^2&-\gamma/2\end{pmatrix}\ , \quad
	K 
		= \begin{pmatrix}-\gamma &0&2\omega\\0&-\gamma&-2\omega \\-\omega&\omega&-\gamma	\end{pmatrix}
	\end{equation}
\end{subequations}
and $\Svec_{\text{inh}}= \gamma \hbar(2\nbar+1)/2\ (1,1,0)^\trans$. The solutions of the time evolution of the first order moments are given by $\ev{\Xvec}\!(t) = \exp(G t) \ev{\Xvec}\!(0)$. For the second order moments we get $\Svec(t)= \exp(K t)\Svec(0)+K^{-1}\qty(\exp(K t)-\mathcal{I}) \Svec_{\text{inh}}$, {where $\mathcal{I}$ denotes the identity operator.}

The two phase-space coordinates $\hat{q}$ and $\hat{p}$ are linked to the annihilation and creation operator $\hat{a}_\omega$ and $\hat{a}^\dagger_\omega$ of the mode by
\begin{subequations}
	\begin{eqnarray}\label{eq:p,q-Def}
	\hat{q}
		&=&\sqrt{\frac{\hbar}{2 M \omega}}\qty(\hat{a}^\dagger_\omega +\hat{a}_\omega),\\
	\hat{p} 
		&=& i \sqrt{\frac{\hbar}{2}M\omega} \qty(\hat{a}^\dagger_\omega -\hat{a}_\omega).
	\end{eqnarray}
\end{subequations}

Summing up, $\omega$, $\gamma$, and $\nbar$ are coded into a state by the dynamics \eqref{eq:ME}, but in addition a state specified initially e.g.~in the Fock basis acquires an $\omega$-dependence due to the $\omega$-dependence of the harmonic oscillator energy eigenstates (oscillator length). 
\subsection{QFI of single-mode Gaussian states}
\textit{Gaussian state.} The Wigner function for an arbitrary density matrix $\rho$ of a continuous variable system with a single degree of freedom (such as a single harmonic oscillator) is defined by \cite{PhysRev.40.749}
\begin{equation}
	W(q,p) 
		= \frac{1}{\pi\hbar} \int_{-\infty}^{\infty} \expp{-2i py/\hbar}\mel{q-y}{\rho}{q+y}\dd{y}.
\end{equation}
By definition, a Gaussian state is a state whose Wigner function is Gaussian. Thus, for a Gaussian state of a single harmonic oscillator (such as a single mode of an electro-magnetical field) the Wigner function takes the general form  \cite{RevModPhys.84.621}
\begin{equation}
	W (q,p) 
		= \frac{P}{\pi} \exp\left[-\frac{1}{2} (\Xvec-\ev{\Xvec})^\trans \Sigmavec^{-1}(X-\ev{\Xvec})\right],
\end{equation}
where $\Xvec = \qty(\hat{q},\hat{p})^\trans$ is the quadrature operator, $\Sigmavec$ is the covariance matrix, $\ev{\ldots}\equiv \tr(\rho \ldots)$ defines the expectation value and $P=\tr\rho^2$ is the purity. For single-mode Gaussian states the purity is completely described by the covariance matrix and is given by \cite{PhysRevA.68.012314}
\begin{equation}
	P 
		= \frac{\hbar}{2 \sqrt{\det(\Sigmavec)}}.
\end{equation}

Next, we recall that a general single-mode Gaussian state $\rho$ can always be represented as a rotated squeezed displaced thermal state $\nu$, \ie \cite{Adam1995,RevModPhys.84.621}
\begin{equation}
	\rho 
		= R(\psi)D(\alpha)S(z) \nu S^\dagger(z) D^\dagger(\alpha) R^\dagger(\psi), \label{eq:Gaussian-State-Representation}
\end{equation}
where $S(z)=\exp[(1/2)\qty(z \hat{a}^{\dagger 2} - z^\ast \hat{a}^2)]$ is the squeezing operator, $R(\psi) = \exp(i \psi \hat{a}^\dagger\hat{a})$ denotes the rotation operator and $D(\alpha) = \exp(\alpha \hat{a}^\dagger - \alpha^\ast \hat{a})$ introduces the displacement operator. By introducing $\Nth=\operatorname{tr}(\nu \hat{a}^\dagger \hat{a})$, which denotes the number of initial thermal photons, and $z=r \expp{i\chi}$ the general Gaussian state can be parametrized by five real parameters $\alpha,\psi,r,\chi,\Nth \in \mathbb{R}$. Note that we keep $\Nth$ and $\bar n$ as independent parameters.

\textit{Quantum Fisher information.} We start from a density operator $\rho_\theta$, which depends on an unknown real scalar parameter $\theta$. To estimate this parameter, $m$ independent measurements with the outcome $\vb{\xi}=\qty(\xi_1,\xi_2,\dots,\xi_M)^\trans$ are taken. From the outcome we construct an estimator $\estimator$.   
For unbiased estimators the sensitivity with which a parameter $\theta$ can be measured has a lower bound, the so-called quantum Cramér-Rao bound (QCRB), given by \cite{PhysRevLett.72.3439, holevo2003statistical, holevo2011probabilistic, hayashi2006quantum}
\begin{equation}\label{eq:QCRB}
	\operatorname{Var} [\estimator] 
		\ge \frac{1}{m \qfi{\rho_\theta}{\theta}},
\end{equation}
where $\qfi{\rho_\theta}{\theta}$ denotes the QFI.
The fidelity, defined by  $\mathcal{F}(\rho_1,\rho_2)  = \{\operatorname{tr}[(\sqrt{\rho_1} \rho_2\sqrt{\rho_1})^{1/2}]\}^2$, for two arbitrary single-mode Gaussian states $\rho_1$ and $\rho_2$ is given by \cite{Scutaru1998}
\begin{widetext}
	\begin{equation}\label{eq:Fidelity-result}
	\mathcal{F}(\rho_1,\rho_2) 
		= \frac{2 \exp[-\frac{1}{2} \qty(\ev{\Xvec_1-\Xvec_2})^ T \qty(\Sigmavec_1 + \Sigmavec_2)^{-1}\ev{\Xvec_1-\Xvec_2}]}{\sqrt{\abs{\Sigmavec_1+\Sigmavec_2} + \qty(1-\abs{\Sigmavec_1})\qty(1-\abs{\Sigmavec_2})}-\sqrt{\qty(1-\abs{\Sigmavec_1})\qty(1-\abs{\Sigmavec_2})}} .
	\end{equation}
\end{widetext}
This formula is valid generally for two Gaussian Wigner functions, regardless of the underlying physical system. It remains therefore valid if the two Wigner functions represent states of two different harmonic oscillators, notably harmonic oscillators that can differ in frequency. Using further the fact that the fidelity is linked to the QFI through \cite{PhysRevA.88.040102}
\begin{equation}\label{eq:QFI-Fidelity}
	\qfi{\rho_\theta}{\theta}
		= -2 \eval{\pdv[2]{\mathcal{F}(\rho_\theta, \rho_{\theta+\varepsilon})}{\varepsilon}}_{\varepsilon=0}
\end{equation}
one obtains the general QFI for Gaussian states of a single harmonic oscillator of fixed frequency \cite{PhysRevA.88.040102}
\begin{eqnarray}
	\qfi{\rho_\theta}{\theta} 
		&=& \frac{1}{2} \frac{\tr\qty[\qty(\Sigmavec^{-1} \partial_\theta\Sigmavec)^2]}{1+P^2} + 2 \frac{\qty(\partial_\theta P)^2}{1- P^4} \nonumber\\ 
		& &+ \qty(\partial_\theta\ev{\Xvec})^\trans \Sigmavec^{-1}\partial_\theta\ev{\Xvec}. \label{eq:General-QFI}
\end{eqnarray}
By following the approach adopted by Jiang in Ref. \cite{PhysRevA.89.032128} the same result can be obtained \cite{PhysRevA.97.012125}.

\section{Undamped case}
This section provides a scheme for the calculation of and results for the QFI relevant for estimating the frequency $\omega$ in the case of no damping. 

\subsection{Scheme for the estimation of the quantum Fisher information\label{SSec:Scheme}}
Firstly we will illustrate that by directly using \eq \eqref{eq:General-QFI} for a frequency measurement one does not get the full QFI, instead one obtains just that part that corresponds to having $\hat{a}^\dagger,\hat{a}$ as frequency-independent generator of the time evolution. For this purpose, we use the known results of the QFI for pure states, where one does not get the full QFI if taking $\hat{a}^\dagger,\hat{a}$ independent of $\omega$. In particular, this means that directly inserting the solution of the dynamics into equation \eq \eqref{eq:General-QFI} will not provide the correct result, as the $\omega$-dependence of the Fock basis is not considered. Lastly, we justify that one can still use \eq \eqref{eq:General-QFI} if one treats the squeezing due to frequency change correctly, which leads to the scheme we propose. 

We consider the case that only the dynamics of the state, and not the initial pure state $\rho_0=\dyad{\psi_0}$ itself, depends on the frequency $\omega$ to be measured. For given Hamiltonian $\mathcal{H} = \hbar\omega \qty(\hat{a}^\dagger \hat{a}+1/2)$, the dynamics of the system is described by $\rho(t)=U(t) \rho (0) U^\dagger(t)$, where $U(t) = \exp(-i t \mathcal{H}/\hbar)$ is the time evolution operator. By neglecting the $\omega$-dependence of $a,a^\dagger$ the QFI is given by \cite{paris2009quantum}
\begin{equation}
	\QFI\qty(\rho_\omega(t);\omega)
		= 4 \operatorname{Var}\qty[t\qty(\hat{a}^\dagger \hat{a}+1/2),\ket{\psi_0}],
\end{equation}
where $\operatorname{Var}\qty[A,\ket{\psi_0}] \equiv \ev{A^2}{\psi_0}-\ev{A}{\psi_0}^2$ denotes the variance.
For a general pure Gaussian state in the form of \eq \eqref{eq:Gaussian-State-Representation}, \ie $\ket{\psi_0}=R\qty(\psi) D\qty(\alpha) S\qty(r\expp{i \chi})\ket{0}$, the QFI reads
\begin{eqnarray}\label{eq:Scheme-Motivation}
	\QFI\qty(\rho_\omega(t);\omega) 
		&=& 4 \alpha^2t^2\qty[\cosh(2r)+\cos(\chi)\sinh(2r)]\nonumber\\
		& &+2t^2\sinh[2](2r).
\end{eqnarray}
Next, we determine the same QFI by directly using equation \eqref{eq:General-QFI}. For this we first use that we can write the time-evolved density operator in the following way:
\begin{equation}
	\rho_\omega(t) 
		=  R(\zeta)D(\alpha)S(z) \dyad{0} S^\dagger(z) D^\dagger(\alpha) R^\dagger(\zeta),
\end{equation}
where $\zeta= \psi-\omega t$. Using 
$\sigma=e^{-r}$, 
equation (16) from  \cite{PhysRevA.88.040102} can be rewritten as \footnote{after correcting $r\to-r$ and $\chi\to2\chi$ in the definitions of \cite{PhysRevA.88.040102}.}
\begin{eqnarray}
	\QFI\qty(\rho_\omega(t);\zeta) 
		&=& 4 \alpha^2\qty[\cosh(2r)+\cos(\chi)\sinh(2r)]\nonumber\\
		& &+2\sinh[2](2r).
\end{eqnarray}
Thus, with $\dd/\dd{\omega}=-t\ \dd/\dd{\xi}$ we get the same result
as obtained in equation \eqref{eq:Scheme-Motivation}, 
of which we have demonstrated that by directly using equation \eqref{eq:General-QFI} the $\omega$-dependence of the basis is not considered. 

In order to consider all frequency dependencies correctly, we have developed the following scheme for the estimation of the QFI: 
\begin{enumerate}
	\item Start with an initial Gaussian state given in the Fock basis $\qty{\ket{n}_{\omega_0}}$.
	\item Perform a sudden change of frequency $\omega_0\to\omega$, which corresponds to a squeezing, at time $t=0$.
	\item Evolve the quantum state with respect to the new frequency $\omega$.
	\item Estimate the QFI $\qfi{\rho_{\omega} (t)}{\omega}$ by using \eq \eqref{eq:General-QFI}.
	\item Take the limit $\omega\to\omega_0$.
\end{enumerate}
The sudden change of frequency $\omega_0\to\omega$ at time $t=0$ ensures that also the frequency dependence of the basis is considered. Furthermore, it can be shown that the frequency jump corresponds to squeezing (\sApp \ref{App:Change-of-Basis}), \ie
\begin{equation}
\ketf{n}{\omega_0} 
= S_{\omega}\qty(s) \ketf{n}{\omega},
\end{equation}
where $s=- \tanh[-1](y_1)$ and $y_1 = \qty({\omega_0}-\omega)/\qty({\omega_0}+\omega)$.

It should be noted that the introduced scheme is only needed to
determine the QFI for a frequency measurement using  \eq
\eqref{eq:General-QFI}. For pure states, for example, the QFI can be
determined directly 
from the overlaps of the states propagated with slightly different
frequency \cite{braun2011ultimate}, or, equivalently, 
from the variance of the local generator, taking into account the
$\omega$-dependence of $\hat{a}_\omega,\hat{a}^\dagger_\omega$ (\sApp
\ref{App:QFI-Pure-State}). Furthermore, it should be noted that since
the Fock basis does not depend on the damping constant, the introduced
scheme is not needed for calculating the QFI for the estimation of
$\gamma$.

\subsection{Result for QFI for vanishing damping}
By using the introduced scheme we now
determine the QFI for the estimation of $\omega$ for the general
Gaussian state given in \eq \eqref{eq:Gaussian-State-Representation}. 
For a time evolution of the Gaussian state with the harmonic
oscillator $\mathcal{H} 
= \hbar\omega \qty(\hat{a}^\dagger_\omega \hat{a}_\omega+1/2)$ follows the result (\sApp \ref{App:QFI-Calculation})
\begin{widetext}
	\begin{eqnarray}\label{eq:QFI-undamped}
	\omega^2 \qfi{\rho(\tau)}{\omega}
		&=& C_3 + 2 C_1 \sin[2](\tau)\bigg[\sinh[2]( 2r)\cos[2](\chi+2\psi-\tau) + 1+ 2C_2\alpha^2\qty(\cosh( 2r) + \cos(\chi+4\psi-2\tau)\sinh( 2r))\bigg]\nonumber\\
		& &+2C_1\tau\sin(\tau)
                    \bigg[4C_2\alpha^2\cos(2\psi-\tau)\cosh (2r) +
                    \cos(\chi+2\psi-\tau)\qty(4C_2\alpha^2 \sinh (2r)
                    + \sinh (4r))\bigg]\nonumber\\ 
		& &+ 2 C_1 \tau^2 \qty[2C_2\alpha^2\qty(\cosh(
                    2r)+\cos \chi\sinh (2r))+\sinh[2](2r)], 
	\end{eqnarray}	
\end{widetext}
where $\tau=\omega t$ and 
 \begin{subequations}
 \begin{eqnarray}
 	C_1 &=& \frac{(1+2\Nth)^2}{1+2\Nth(1+\Nth)},\\
 	C_2 &=& \frac{1}{C_1(1+2\Nth)},\\
 	C_3 &=& \Nth (1+\Nth)\qty[\ln(\frac{1+\Nth}{\Nth})]^2.
 \end{eqnarray}
 \end{subequations}
The {first term ($C_3$) of \eq \eqref{eq:QFI-undamped}} results from the $\omega$-dependence of
the initial photon number $\Nth$, the second term 
is due to the
$\omega$-dependence of the Fock basis, and the term $\propto t^2$
arises from 
$\hat{a}^\dagger, \hat{a}$ as generator of the time evolution. 

For an initial thermal state $\rho(0)=\nu$, 
\eq \eqref{eq:QFI-undamped} reduces to
\begin{equation}\label{eq:QFI-Thermal-state}
	\qfi{\nu(\tau)}{\omega} 
		= \frac{2 C_1 \sin^2(\tau)+C_3}{\omega^2}.
\end{equation}
Thus, a measurement with $t>\pi/2\omega$ does not provide any additional information regarding the frequency and the QFI has an upper bound $(2 C_1+C_3) /\omega^2$---where $C_1$ itself is bounded by $C_1 \in [1,2]\ \forall \Nth$ and $C_3$ is bounded by $C_3 \in [0,1]\ \forall \Nth$. 
Furthermore, the result demonstrates that one can measure the frequency of a mode of an e.m. field without any light at all, just from the vacuum fluctuations. The latter have been measured directly in \cite{Riek420}. 

While our results from \eq \eqref{eq:QFI-undamped} agree with the obtained QFI for a coherent state \cite{braun2011ultimate}, our result in \eq \eqref{eq:QFI-Thermal-state} contains an extra term $C_3/\omega^2$ due to the consideration of the $\omega$-dependence of $\Nth$ neglected in \cite{braun2011ultimate}. It should also be noted that our result agrees with the result by calculating the QFI via the variance in the case of a general pure Gaussian state (\sApp \ref{App:QFI-Pure-State}).

\textit{Optimal state}. The QFI can be drastically increased by
displacing and/or squeezing the initial thermal state. In both cases,
the QFI acquires a part proportional to $t^2$ that always dominates at sufficiently large times. For an initial 
state displaced with $\alpha\in\mathbb R$,
the part proportional to $t^2$ has its maximum at $\chi=0$. 
We further point out that the long-term behavior of the QFI for a
squeezed thermal state also improves 
due to additional displacing.

The optimal choice of thermal photons $\Nth$ depends on the initial
state. 
If the QFI is dominated by the term{s} due to the squeezing, a
high number of photons is favorable. If, on the other hand, the term{s}
due to the displacement{, which are $\propto (1+2\Nth)^{-1}$, dominate}, the lowest possible number of
photons is desirable. The behavior can be well explained by the
Wigner function. A larger $\Nth$ is equivalent to a wider distribution
of the state. This means that a small shift in the Wigner function of
the displaced state, \eg due to the time evolution, is less measurable
for larger $\Nth$.  Consequently, the enlargement of the thermal
photons counteracts the additional gain of the displacement. The benefits of squeezing, 
on the other hand, increase with the thermal photon number.  This can be directly
  from eq.\eqref{eq:QFI-undamped}
  seen, since its QFI is proportional to $C_1$, which is also the only
  term that increases with  $\Nth$. 
\section{Damped case}

In this section we will calculate the QFI for mixed Gaussian states
for the damped harmonic oscillator for estimating the oscillator frequency and damping
constant. Furthermore, we determine the optimal measuring scheme and
the optimal measuring time and we demonstrate that with existing carbon nanotube resonators it should be possible {to achieve a mass sensitivity of the order of an electron mass $\si{\hertz^{-1/2}}$.}
\subsection{Measuring the oscillator frequency}
By sticking to the scheme explained in Sec.\ref{SSec:Scheme}{,} we obtain the exact expression for the QFI 
for a general initial Gaussian state by considering the time evolution
{given by the } 
ME \eqref{eq:ME}, which can be found in the Appendix, in \eq
\eqref{eq:App-QFI-3parts} to \eqref{eq:App-QFI-abbr}. 
However, since the solution is too heavy to report here, we will first
look at the long-term behavior and then limit ourselves to specific
initial states---coherent state and squeezed state. 
\subsubsection{Long-term behavior \label{SSSec: Long-Term Behaviour}}
For longer periods, the solution of ME \eqref{eq:ME} relaxes to the
thermal equilibrium state, \ie for $t\gg {\gamma^{-1}}$, 
\begin{equation}
	\rho \xrightarrow{t\gg \gamma^{-1}} 
		\expp{-\hbar\omega/k_\textsc{B} T
                }/\tr(\expp{-\hbar\omega/k_\textsc{B} T})\equiv
                \rho_\infty. 
\end{equation}
It should be remembered that the thermal equilibrium state as well as the mean thermal photon number $\nbar$ also depend on the oscillator frequency $\omega$ itself. It can therefore be expected that the QFI does not vanish due to the dependency of the final state on the frequency. Since both first order moments vanish, \ie $\lim_{t\to\infty}\ev{\Xvec}=0$, only the first two terms of \eq \eqref{eq:General-QFI} contribute to QFI and calculation yields
\begin{eqnarray}	
	\qfi{\rho_\infty}{\omega} 
		&=& \frac{1}{2\omega^2}\Bigg[2\nbar(1+\nbar)\ln^2\!\left(\frac{1+\nbar}{\nbar}\right) \nonumber\\
		& & \phantom{ \frac{1}{2\omega^2}\Bigg[}+ \frac{1+4 \nbar(1+\nbar)}{1+2 \nbar(1+\nbar)}\Bigg].\label{eq:QFI-Long-Term-ME}
\end{eqnarray}

\begin{figure}[t]
	\includegraphics[width=0.45 \textwidth]{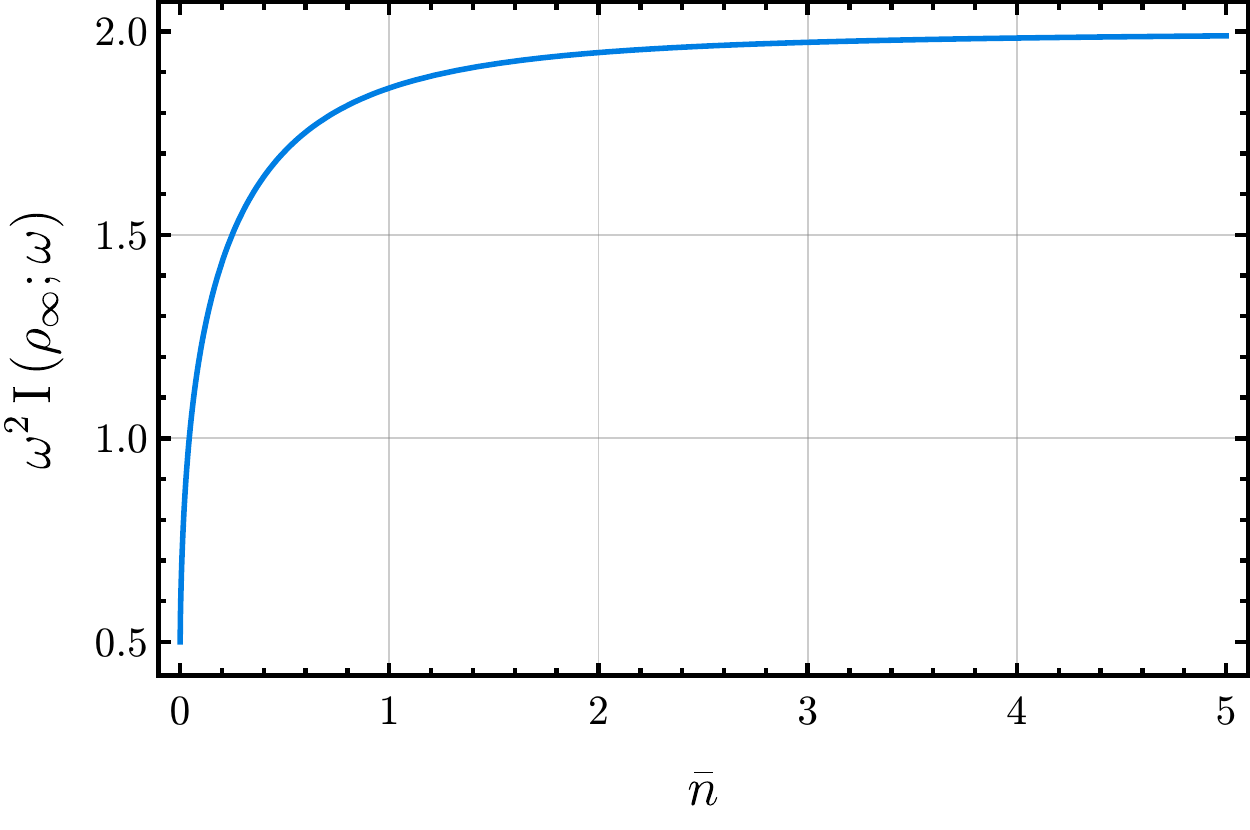}
	\caption{\label{fig:QFI-Long-Term} The long-term behavior of the dimensionless QFI, $\omega^2 \qfi{\rho_\infty}{\omega} $, for a damped Gaussian state for measuring $\omega$ is shown as function of the {thermal photon number of the bath}. In the limit of validity of \eqref{eq:ME}, the result is independent of the damping constant.}
\end{figure}
\begin{figure}[t]
	\includegraphics[width=0.45 \textwidth]{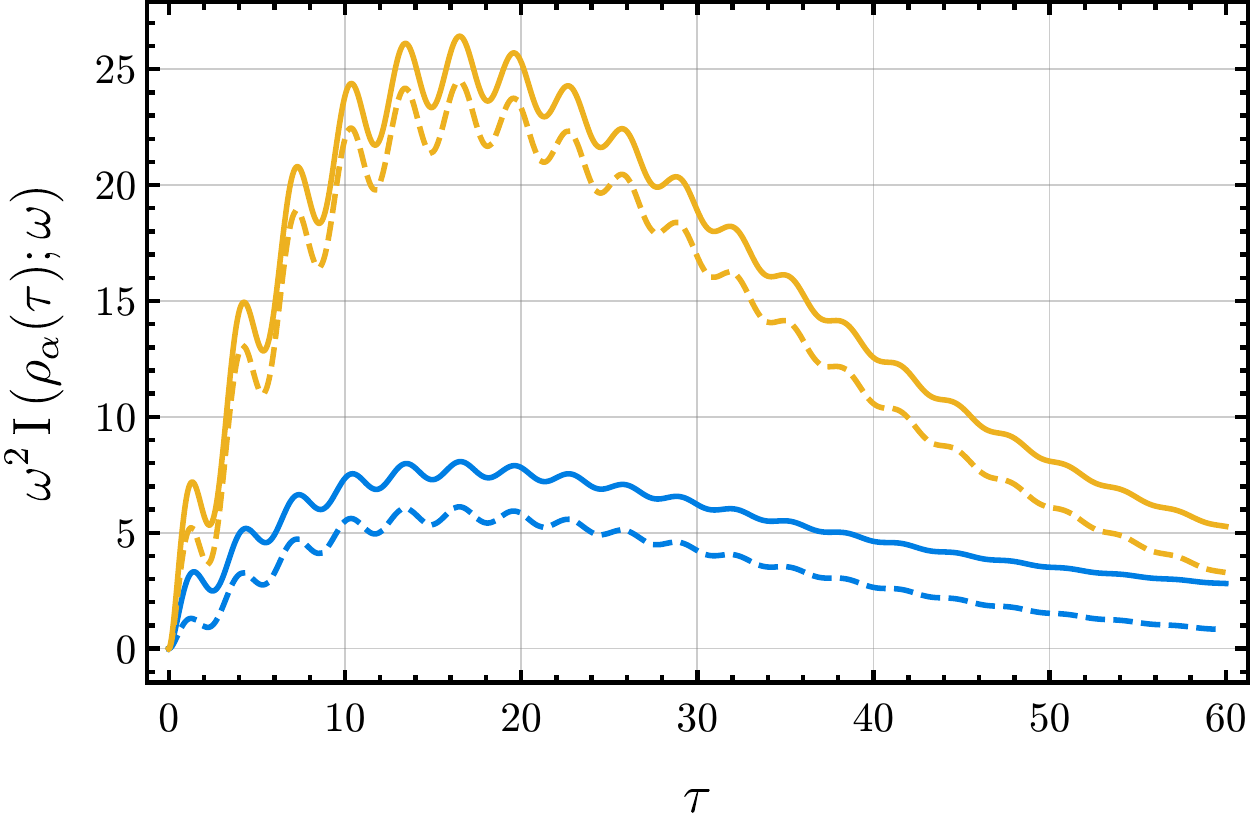}
	\caption{\label{fig:QFI-CS}	The dimensionless QFI,
          $\omega^2 \qfi{\rho_\alpha}{\omega}$, for an initial
          coherent state (solid) is compared with the lower bound $
          \omega^2	\QFI_\alpha	(\tau)$ (dashed) {(see \eq
            \eqref{eq:QFI-Coherent-State-Notation})} for measuring
          $\omega$. Results are depicted for $\nbar=5$ and $g=0.1$:
          blue, $\alpha=1/2$; orange, $\alpha=1$.}	 
\end{figure}
This means that for 
large times, the QFI has an upper bound given by $2/\omega^2$ (see
Fig. \ref{fig:QFI-Long-Term}). The upper bound can be reached in the
high temperature limit. 
As a consequence, a longer measurement does not necessarily yield a
better result for the experiment. In other words, there is an optimal
measurement (OMT) time in which the frequency can be measured best,
which is in accordance with the physical expectations.

\subsubsection{Optimal measurement time and maximal quantum Fisher information}
\textit{Coherent state}. {We start {by} considering an initial coherent state $\rho_\alpha(0)=D(\alpha)\dyad{0}{0}D^\dagger(\alpha)$}. Recall, displacing the initial state is one of the possibilities to strongly increase the QFI in the undamped case. Since displacing the ground state only affects the expectation values of the quadrature operators 
and not the covariance matrix, the QFI of the coherent state can be written as  
\begin{equation}\label{eq:QFI-Coherent-State-Notation}
	\qfi{\rho_\alpha(\tau)}{\omega}
		= \qfi{\rho_0(\tau)}{\omega} +  \QFI_\alpha(\tau),
\end{equation}
where {$\rho_0(\tau)$ 
 denotes the time-evolved ground state and} $\QFI_\alpha(\tau) =
\qty(\partial_\omega\ev{\Xvec})^\trans
\Sigmavec^{-1}\partial_\omega\ev{\Xvec}$.
{The QFI of the ground state is bounded by $2.135/\omega^2$ (\sApp  \ref{App:QFI-Calculation}). Thus, the upper bound of the QFI for the ground state is increased by introducing the system-bath coupling, which can be explained by the $\omega$-dependence of $\nbar$. I.e.~also in the damped case, the {frequency can be measured when the system is initially prepared in the ground state.} 
Straightforward calculation leads to
\begin{equation}\label{eq:QFI-CS-Damped}
	\QFI_\alpha	(\tau)	
		= \frac{4 \alpha ^2}{\omega ^2} \frac{\sin ^2(\tau )+\tau\sin (2 \tau ) + \tau^2}{(2 \nbar+1) e^{g \tau }-2 \nbar},
\end{equation}
where $g=\gamma/\omega$ introduces a dimensionless damping constant. Thus, for frequency measurements an as big as possible displacement is recommended.

Since the QFI of the ground state is {bounded (and small)},
$\qfi{\rho_\alpha(\tau)}{\omega} \approx \QFI_\alpha(\tau)$ applies {for $\alpha^2\gg g^2 \nbar$ (by assuming $n\gg1$ and $g\ll1$).}
For high enough temperatures{, $\nbar \gg \alpha^2/g^2$,}  $\QFI_\alpha(\tau)$ becomes arbitrarily small and the QFI is then described by the QFI of the ground state. In other words, displacement only improves frequency measurements for resulting mean energies larger than the thermal energy.

\begin{figure}[t]
	\includegraphics[width=0.45 \textwidth]{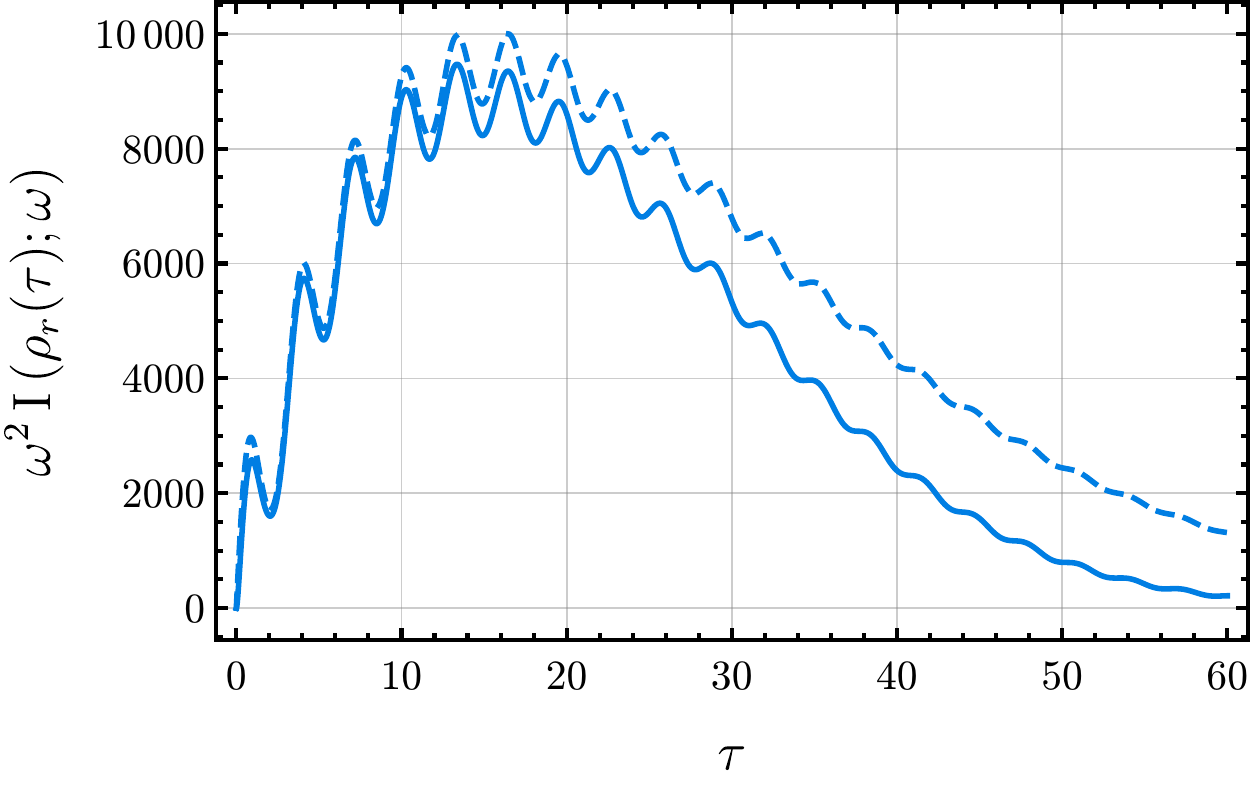}
	\caption{\label{fig:QFI-SS} The dimensionless QFI, $\omega^2 \qfi{\rho_r}{\omega}$, for an initial squeezed state  (solid) is compared with the approximation from \eq \eqref{eq:QFI-SS-approx} (dashed) for measuring $\omega$. Results are depicted for $\nbar=0.01$, $g=0.1$ and $r=2.5$.} 
\end{figure}
By neglecting small oscillations, maximization of \eq \eqref{eq:QFI-CS-Damped} provides the maximal QFI $\QFI_\text{max}(\rho,\theta)	\equiv \max_{\tau} \QFI(\rho,\theta)$ and the optimal measurement time $\tau_\text{max}$ with $\QFI(\rho(\tau_\text{max}),\theta)=\QFI_\text{max}(\rho,\theta)$, \ie 
\begin{eqnarray}
	\QFI_\text{max}(\rho_\alpha(\tau),\omega) 
		&=& -\frac{2\alpha^2}{\nbar g^2\omega^2 }\mathcal{W}\qty(-\frac{4\nbar}{\expp{2}(1+2\nbar)})\nonumber\\
		& &\times \qty[2+\mathcal{W}\qty(-\frac{4\nbar}{\expp{2}(1+2\nbar)})],\\
	\tau_\text{max} 
		&=& \frac{1}{g}\qty[2+\mathcal{W}\qty(-\frac{4\nbar}{\expp{2}(1+2\nbar)})],
\end{eqnarray}
where $\mathcal{W}(z)$ denotes the Lambert $\mathcal{W}$ function defined by $z=\mathcal{W}(z) \expp{\mathcal{W}(z)}, z\in\mathbb{C}$. 

{The Taylor series for $ \qfi{\rho_0(\tau)}{\omega}$ at $\nbar \gg1$ is 
\begin{equation}
	\qfi{\rho_0(\tau)}{\omega} 
		=\frac{2}{\omega^2}\qty[1-\frac{\cos[2](\tau)}{(\expp{g\tau}-1)\nbar}]+\order{1/\nbar^2}.
\end{equation}
That means that for high temperatures $\bar n\gg 1$ the QFI of the ground state decays 
  faster ($\sim \expp{-g\tau}$) than $\QFI_\alpha(\tau)$ ($\sim \tau^2\expp{-g\tau}$) 
and varies only slightly close to the time $\tau_\text{max} $. Consequently, the use of $\QFI_\alpha (\tau)$ 
for estimating the optimal measurement time leads, even in this range, 
to a good result of 
the OMT (see Fig. \ref{fig:QFI-CS}).}
Furthermore, it should be noted that the smaller $g$, the larger $\nbar$ can be, so that the OMT is still very well described by $\QFI_\alpha (\tau)$.
By reducing the system-bath coupling, the maximal 
QFI increases proportionally to $\propto g^{-2}$. However, it should be noted that the OMT also 
increases proportionally to $\propto g^{-1}$. 

Thus, it is a natural to consider time as a resource and to introduce the rescaled maximal QFI $\QFI_\text{max}^{(t)}(\rho,\theta)	\equiv \max_{t} \QFI(\rho,\theta)/t$ and the optimal measurement time $\tau_\text{max}^{(t)}$ that maximizes it. 
For an initial coherent state we get
\begin{eqnarray}
	\QFI_\text{max}^{(t)}(\rho_\alpha(t),\omega)
		&=&- \frac{2\alpha^2}{\nbar g \omega}  \mathcal{W}\qty(-\frac{2\nbar}{\expp{}(1+2\nbar)}),\\
	\tau_\text{max}^{(t)} 
		&=&\frac{1}{g}\qty[1+\mathcal{W}\qty(-\frac{2\nbar}{\expp{}(1+2\nbar)})].
\end{eqnarray}
Taking time into account as a resource leads to a reduction of the OMT. 

\textit{Squeezed state}.
Besides displacement, squeezing the initial state is another possibility to increase the QFI in the undamped case. Therefore, we determine the QFI for a squeezed state $\rho_r(0) = S(r)\dyad{0}{0}S^\dagger(r)$. 
For the coherent state we have seen that reducing the temperature leads to an increase in the QFI. This behavior is reasonable, since 
increased temperature implies increased damping according to the master equation \eqref{eq:ME}.  
A similar behavior can be observed here with the squeezed state. The QFI for {a vanishing bath temperature, \ie $\nbar=0$,} reads
\begin{widetext}
\begin{eqnarray}
	\QFI(\rho_r (t),\omega) 
		&=& \left[8 \omega ^2 \left(2 e^{g \tau } \sinh ^2(r)+e^{2 g \tau }-\cosh (2 r)+1\right)\right]^{-1}\nonumber\\		
		& & \times\Big[16 \tau  \sinh (2 r) \sin (2 \tau ) \left(e^{g \tau }+\cosh (2 r)-1\right)-4 \left(e^{g \tau }-1\right) \cosh (2 r) (2 \cos(2 \tau )-3)\nonumber\\
		& &\phantom{\times\Big\{}+4 e^{g \tau } \left(e^{g \tau }-1\right)+\left(8 \tau ^2+1\right) \cosh (4 r)-8 \sinh ^2(r) \cosh ^2(r) \cos (4 \tau )-8 \tau ^2-8 \cos (2 \tau )+7\Big].
\end{eqnarray}
\end{widetext}
Alternatively, for high squeezing and low temperatures, \ie $r\gg1$ and $\nbar\ll1$, the QFI can be approximated as (see Fig. \ref{fig:QFI-SS})
\begin{eqnarray}\label{eq:QFI-SS-approx}
	\QFI(\rho_r (\tau),\omega) 
		&\approx& \frac{\expp{2r} \qty[2\tau + \sin(2\tau)]^2}{4 \omega^2 \qty(\expp{g \tau}-1)\qty(1+2\nbar)}.
\end{eqnarray}
Thus, the QFI can be significantly increased by squeezing also for an initial thermal state.
Neglecting the oscillations, the OMT can be determined to
\begin{equation}
	\tau_\text{max} 
		= \frac{1}{g}\qty[2+\mathcal{W}\qty(-2/\expp{2})] \approx \frac{1.59}{g}.
\end{equation}
For sufficiently high squeezing and low temperature, the OMT does not depend on the squeezing and temperature anymore.
\subsection{Measuring the damping constant}
Next we consider the QFI for the 
estimation of the damping
constant. First of all, the QFI disappears
for large times, i.e. $\QFI(\rho_\infty,\gamma)=0$. This can be seen
directly from the fact that the final thermal state (for the master
equation approach) itself no longer depends on the damping
constant. In other words, there is again an OMT.  

After a straightforward but long and tedious calculation we find for the QFI of a general Gaussian state
\begin{widetext}
\begin{eqnarray}
	\qfi{\rho(\tau)}{\gamma} 
		&=& \frac{P^2(\tau)g^2\tau^2}{\gamma^2 \expp{4 g\tau}} \Bigg\{ \alpha^2\expp{2g\tau}  \qty[A_1 \qty(\cosh(2r)-\cos(\chi) \sinh(2r))+a_{1,\tau}]
			+\frac{2P^4(\tau)}{1-P^4(\tau)}\qty[A_1^2 + A_1 \qty(a_{1,\tau}-a_1) \cosh(2r)-a_1 a_{1,\tau}]^2\nonumber\\
		& &\phantom{\frac{P^2(\tau)\tau^2}{\omega^2 \expp{4 g\tau}} \Bigg\{}+\frac{P^2(\tau)}{1+P^2(\tau)} \bigg[A	_1^4+A_1^2 \left(a_1^2+a_{1,\tau}^2\right) \cosh(4 r)+2 A_1 (a_{1,\tau}-a_1)  \left(A_1^2-a_1 a_{1,\tau}\right)\cosh (2 r)\nonumber\\
		& &\phantom{\frac{P^2(\tau)\tau^2}{\omega^2 \expp{4 g\tau}} \Bigg\{+\frac{P^2(\tau)}{1+P^2(\tau)} \bigg[}-4 a_1  a_{1,\tau}A_1^2+a_1^2 a_{1,\tau}^2\bigg]\Bigg\},
\end{eqnarray}
\end{widetext}
where $	a_1 = 1+2\nbar$, $a_{1,\tau} = (\expp{g \tau}-1)a_1$, $A_1= 1+2\Nth $ and
\begin{equation}
	P(\tau) = e^{g\tau} \qty[A_1^2+a_{1,\tau}^2 +2 a_{1,\tau} A_1 \cosh(2r)]^{-1/2}.
\end{equation}
The result does not depend on the rotation angle $\psi$, but only on the squeezing angle $\chi$. In contrast to frequency measurement, the QFI for measuring $\gamma$ is maximized for $\chi=\pi$. This is in agreement with the physical expectation, as the relevant dynamic here is the relaxation of $\ev{\Xvec}$. To illustrate the result, we again consider specific initial states --- thermal state, displaced thermal state and squeezed state.

\textit{Thermal state}.
The QFI of a thermal state $\nu$ can be written as
\begin{equation}
	\QFI(\nu(\tau),\gamma) 
		=\frac{(\nbar-\Nth)^2 g^2\tau^2}{\gamma^2\qty[ \qty(\expp{g\tau} - 1)\nbar + \Nth ]\qty(\expp{g \tau} \qty(1+\nbar) + \Nth-\nbar)}.
\end{equation}
The greater the deviation of the initial temperature from the bath temperature, the better $\gamma$ can be measured. In particular, for a vanishing deviation, i.e. $\Nth=\nbar$, the QFI vanishes, since in this case the state has no dynamics at all. For $\nbar=0$, the OMT is given by
\begin{equation}
	\tau_\text{max} = \frac{2+\mathcal{W}\qty(2\Nth\expp{-2})}{g}.
\end{equation}

\textit{Displaced thermal state}.
For an initial displaced thermal state $\rho_{\alpha,\Nth}(0)= D(\alpha)\nu D^\dagger(\alpha)$ the QFI for measuring $\gamma$ reads
\begin{equation}
	\QFI(\rho_{\alpha,\Nth}(\tau),\gamma) 
		=\QFI(\nu(\tau),\gamma) + \frac{\alpha^2 g^2\tau^2}{\gamma^2\qty[2\Nth -2\nbar +\expp{g \tau}(1+2\nbar)]}.
\end{equation}
Particularly for $\Nth=\nbar$, the QFI simplifies to
\begin{equation}
	\QFI(\rho_{\alpha,\nbar}(\tau),\gamma) 
		= \frac{\alpha^2 g^2\tau^2}{\gamma^2\expp{g\tau}(1+2\nbar)}
\end{equation}
and the OMT is given by $\tau_\text{max}=2/g$.  For $\Nth=\nbar$ only the 3\textsuperscript{rd} part of equation \eqref{eq:General-QFI} contributes to the QFI, i.e. the QFI results solely from the relaxation of $\ev{p},\ev{q}$. By considering the rescaled QFI the OMT reduces to $\tau_\text{max}^{(t)}=1 /g$.

\textit{Squeezed state}.
The low temperature limit behavior, \ie $\nbar=0$, of the QFI for an initial squeezed state $\rho_r$ is given by
\begin{equation}
	\qfi{\rho_r(\tau)}{\gamma} 
		= \frac{\qty[\expp{2 g \tau}-2\qty(\expp{g \tau}-1)]g^2\tau^2 \sinh^2(r)}{\gamma^2\qty(\expp{g \tau}-1)\qty[2(\expp{g\tau}-1)\sinh^2 (r)+\expp{2g\tau}]}.
\end{equation}

The sensitivity with which the damping parameter can be measured improves by squeezing, displacing and/or a temperature deviation (see Fig. \ref{fig:QFI-gamma}).

\begin{figure}[b]
	\includegraphics[width=0.45 \textwidth]{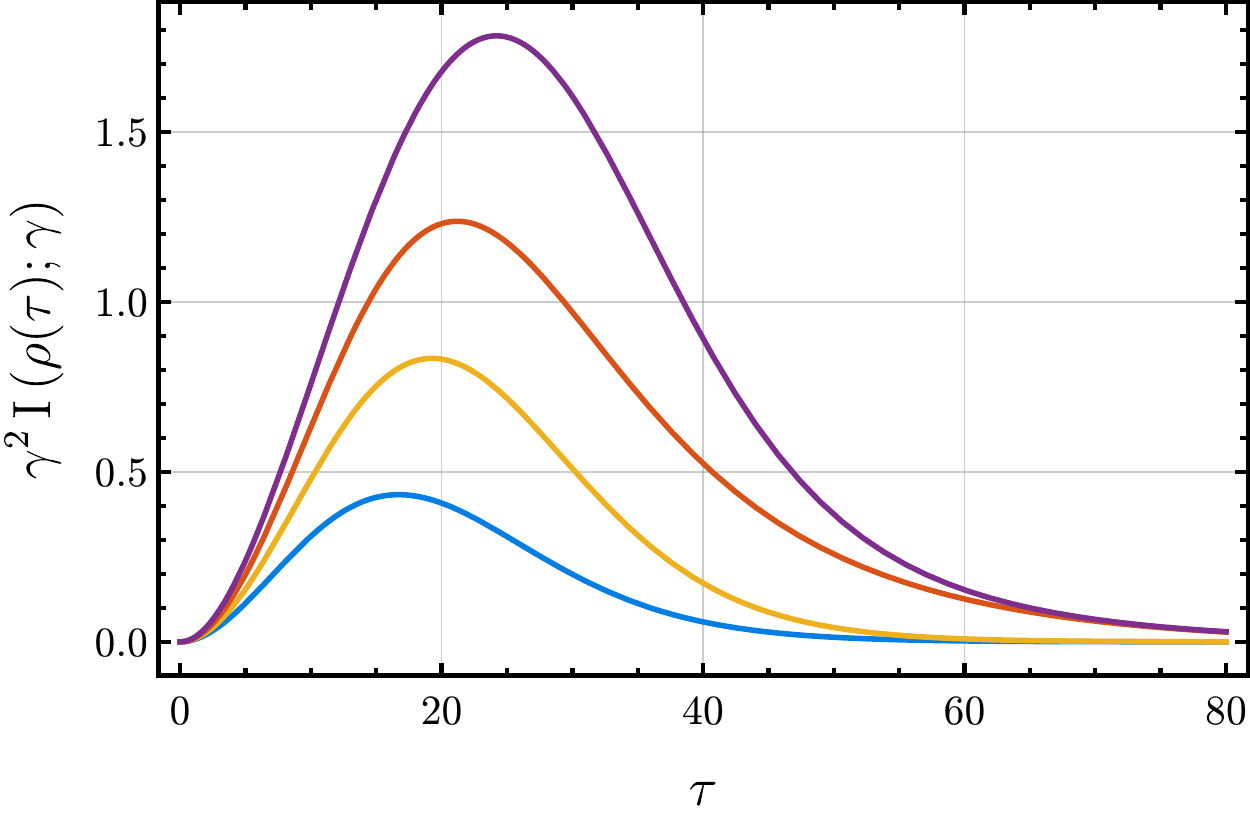}
	\caption{\label{fig:QFI-gamma} The dimensionless QFI, $\gamma^2 \qfi{\rho}{\gamma}$, for measuring $\gamma$  is shown. Results are depicted for $\nbar=1$, $g=0.1$, $\chi=\pi$ and: blue, $\Nth=5$, $\alpha=r=0$; orange, $\Nth=10$, $\alpha=r=0$; red, $\Nth=10$, $\alpha=1$, $r=0$; purple, $\Nth=10$, $\alpha=1$, $r=1/2$.}
\end{figure}

\subsection{Nano-mechanical resonators}
In the following we apply the results obtained to nano-mechanical resonators, which function as precision mass sensors as their resonance frequency changes when additional mass is adsorbed. More precisely, we consider carbon nanotube resonators. Using the QCRB \eqref{eq:QCRB} 
and $\omega = \sqrt{D/M}$, where $D$ is the effective spring constant of the harmonic oscillator, the smallest $\delta M$ that can be resolved from $m$ measurements of the resonance frequency is given by
\begin{equation}\label{eq:delta_M_min}
	\delta M_\text{min} = \frac{2 M}{ \omega \sqrt{m \QFI_\text{max}(\rho,\omega)}}.
\end{equation}
Assuming a coherent state with oscillation amplitude of about $\SI{10}{\nano\meter}$ for the carbon nanotube resonator in \cite{Chaste2012} ($M=\SI{3d-22}{\kilogram}$, $\omega = 2\pi \times \SI{1.865}{\giga\hertz}$, $T=\SI{4}{\kelvin}$ and {$Q\sim \num{d3}$}), $\delta M_\text{min}$ according to \eqref{eq:delta_M_min} is 
slightly below one proton mass. Using the OMT given by $t_\text{max}=\SI{270}{\nano \second}$, the sensitivity corresponds to $\delta M_\text{min} \sqrt{t_\text{max}}=0.8\, m_e \si{\hertz^{-1/2}}$, which is less than $1/4000$ of the  experimentally determined mass sensitivity of slightly more than one proton mass after $\SI{2}{\second}$ averaging time.

In \cite{braun2011ultimate} the theoretically achievable $\delta M_\text{min}$ for the carbon nanotube resonator in \cite{Jensen2008} ($M=\SI{d-21}{\kilogram}$, $\omega = 2\pi \times \SI{328.5}{\mega\hertz}$, $T=\SI{300}{\kelvin}$ and $Q\sim \num{d3}$) was determined to the order of a thousandth of an electron mass. 
{Including the system-bath coupling, $\delta M_\text{min}$ increases to about $74$ proton masses, where the OMT is given by $t_\text{max}=\SI{1.5}{\micro \second}$. This result is equivalent to $ \SI{0.8}{\atomicmassunit/\sqrt{\hertz}}$, which approximately corresponds to one hundredth  of the $\SI{78}{\atomicmassunit}/ \si{\sqrt{\hertz}}$ achieved in the experiment.}

\section{Conclusions}
In summary, we have derived the quantum Cramér-Rao bound for measuring the oscillator frequency and damping constant encoded in the dynamics of a general mixed single-mode Gaussian state of light, including damping through photon loss described by a Lindblad master equation.
We first demonstrated that the known solution for the QFI for Gaussian
states of a single harmonic oscillator of fixed frequency cannot be
directly applied to frequency {measurements}. Next, we presented a
scheme through which the frequency estimation can nevertheless be
based on the results of Pinel et al. \cite{PhysRevA.88.040102}. 

Furthermore, we have shown that displacing and/or squeezing the
initial state significantly increases the precision with which
$\omega$ and $\gamma$ can be estimated. 
For measuring $\omega$ and $r\neq0$, $\chi=0$ is optimal, whereas for measuring $\gamma$, $\chi=0$ maximizes the QFI.

Our results can serve as important benchmarks for the precision of frequency measurements of any harmonic oscillator with given damping. In particular, we found optimal measurement times that limit the sensitivity per $\sqrt{\text{Hz}}$  with which frequencies can be measured, in contrast to the undamped case, where e.g.~coherent states lead to growing {QFI} for arbitrarily large times. 

\appendix
\section{Change of basis\label{App:Change-of-Basis}}
By presenting the scheme for the estimation of the QFI for measuring $\omega$ we made use of the fact that the frequency jump corresponds to squeezing. Next we prove the statement, \ie the following formula
\begin{equation}\label{eq:Chap5-squeezednumberstate}
	\ketf{n}{\omega_0} 
		= S_{\omega}\qty(s) \ketf{n}{\omega}  ,
\end{equation}
where $s=- \tanh[-1](y_1)$. For the sake of simplicity the two parameters 
\begin{equation}\label{eq:y,q-parameters}
	y_1 = \frac{{\omega_0}-\omega}{{\omega_0}+\omega},\quad 
	y_2 = \frac{2\sqrt{{\omega_0}\omega}}{{\omega_0}+\omega}
\end{equation}
are introduced. A squeezed number state is given by \cite{PhysLetA.229.135}
\begin{eqnarray}
	& &\braf{m}{\omega} S_{\omega} \qty(s) \ketf{n}{\omega} \nonumber\\
		&=& \frac{\sqrt{n!}}{\cosh^{n+1/2}\qty|s|} \sum\limits_{j=0}^{ \lfloor n/2\rfloor} \frac{\qty(-d^\star)^j \cosh^{2j}\qty|s|}{\qty(n-2j)!j!}\nonumber\\
		& &\times\sum\limits_{k=0}^{\infty} \frac{d^k \sqrt{\qty(n-2j+2k)!}}{k!}\underbrace{\braketf{m}{n-2j+2k}{\omega}{\omega}}_{=\delta_{m,n-2j+2k}},
\end{eqnarray}
where $d\equiv\qty(s/2\qty|s|)\tanh\qty|s|$ and $\lfloor n/2\rfloor$ denotes the floor function. With $m=n-2j+2k$ and $k\in \mathbb{N}$ we get $k=j+\frac{m-n}{2}\in\mathbb{N}$. This means in particular that $m$ and $n$ must be both even or both odd numbers, otherwise the overlap disappears. If $m$ and $n$ satisfy this condition and by using $\cosh\qty|s|=1/y_2$ and $d=-y_1/2$, the expression can be further simplified as follows
\begin{eqnarray}
	& &\braf{m}{\omega} S_{\omega} \qty(s) \ketf{n}{\omega}\nonumber\\
		&=& \sqrt{y_2 m! n!}\sum\limits_{j=0}^{\lfloor n/2\rfloor} \frac{(-1)^{j+\frac{m-n}{2}}\qty(\frac{y_1}{2})^{2j+\frac{m-n}{2}}}{\qty(n-2j)!j!\qty(j+\frac{m-n}{2})!} y_2^{n-2j}.
\end{eqnarray}
By changing the index of summation to  $l=n-2j$ we get the new upper bound of $\operatorname{min}\qty(m,n)$, where $\operatorname{min}\qty(m,n)$ denotes the smaller of the two integers $m$, $n$. $l$ is also bounded by $m$, since $k=j-\frac{n-m}{2} = \frac{m-l}{2}\in\mathbb{N}$ and thus $l\le m$. Using the new index of summation we get \cite{smith1969overlap}
\begin{eqnarray}
	& &\braf{m}{\omega} S_{\omega} \qty(s) \ketf{n}{\omega}\nonumber\\
		&=& \sqrt{\frac{y_2 m! n!}{2^{m+n}}}\sum\limits_{l=0,1}^{\operatorname{min}\qty(m,n)} \frac{\qty(2y_2)^l}{l!}\frac{y_1^{(m+n-2l)/2} \qty(-1)^{(m-l)/2}}{\qty(\frac{n-l}{2})!\qty(\frac{m-l}{2})!}\nonumber\\
		&=& \R{\omega}{\omega_0}{m}{n}\nonumber\\
		&=& \braketf{m}{n}{\omega}{\omega_0},
\end{eqnarray}
where $\R{\omega}{\omega_0}{m}{n}$ denotes the overlap matrix element between energy eigenstates of the two oscillators with frequency $\omega$ and $\omega_0$. Since this is true for all $m$, we have proven the formula. Thus, for any density operator follows
\begin{align}
	\rho_{\omega_0}
		=  S_{\omega} \qty(s) \tilde{\rho}_{\omega}S_{\omega}^\dagger \qty(s), 
\end{align}
where $s= -\tanh[-1](y_1)$ and $\tilde{\rho}_\omega$ corresponds to the initial state $\rho_{\omega_0}$ by replacing the frequency $\omega_0$ of the basis with the new frequency $\omega$. Thus, we have shown that the initial frequency change corresponds to a squeezing. It should be noted that in the case of a vanishing frequency change, \ie $\omega_0=\omega$, $y_1=0$, $s=0$ and $S (s=0)=\mathcal{I}$ follow and thus $\rho_{\omega_0}=\rho_{\omega}$ is ensured.
\section{QFI for pure states\label{App:QFI-Pure-State}}
In the following it will be shown that the introduced scheme provides the correct QFI for an undamped pure Gaussian state. Therefore, the QFI is calculated analogously to chapter \ref{SSec:Scheme}, but this time also the $\omega$-dependence of $\hat{a}^\dagger,\hat{a}$ are taken into account. 

This means, we consider the case that only the dynamics of the state, and not the initial state 
\begin{equation}
	\rho_0 = \dyad{\psi_0}, 
\end{equation}
where $\ket{\psi_0}=R\qty(\psi) D\qty(\alpha) S\qty(r\expp{i \chi})\ket{0}$, depends on the frequency $\omega$ to be measured. For given Hamiltonian $\mathcal{H}_\omega = \hbar\omega \qty(\hat{a}^\dagger_\omega \hat{a}_\omega+1/2)$, the dynamics of the system is described by $\rho_\omega=U_\omega \rho_0 U_\omega^\dagger$, where $U_\omega = \exp[-i \omega t  \qty(\hat{a}^\dagger_\omega \hat{a}_\omega+1/2)]$ is the time evolution operator. With the help of the local generator
\begin{equation}\label{eq:LocalGeneratorK}
	\mathscr{K} 
		= i U_\omega^\dagger(t) \pdv{U_\omega(t)}{\omega}
\end{equation}
the QFI can be rewritten as follows \cite{PhysRevLett.98.090401}
\begin{equation}\label{eq:QFI-pure-state}
	\qfi{\rho_\omega}{\omega} 
		= 4 \operatorname{Var}\qty[\mathscr{K},\ket{\psi_0}].
\end{equation}
 If $\vb{A}$ is a Matrix depending on the parameter $x$, $\vb{A}=\vb{A}(x)$, then \cite{snider1964perturbation}
\begin{equation}
	\pdv{x}\expp{\vb{A}(x)} 
		= \qty(\int\limits_{0}^{1} \expp{\alpha \vb{A}(x)}\pdv{\vb{A}(x)}{x}\expp{-\alpha \vb{A}(x)}\dd{\alpha}) \expp{\vb{A}(x)}\!\!.
\end{equation}
Using this formula we can rewrite the local generator $\mathscr{K}$ as \cite{PhysRevA.95.062342}
\begin{equation}
	\mathscr{K} 
		= \frac{t}{\hbar} \int\limits_{-1}^{0} V(\alpha) \pdv{\mathcal{H}_\omega}{\omega}V^\dagger(\alpha) \dd{\alpha},
\end{equation}
where $V(\alpha) = \exp(-i\alpha t \mathcal{H}_\omega/\hbar)$. The derivative of the Hamiltonian $\mathcal{H}_\omega$ with respect to the oscillator frequency $\omega$ reads
\begin{equation}
	\pdv{\mathcal{H}_\omega}{\omega} 
		= \hbar a^\dagger_\omega \hat{a}_{\omega} + \frac{\hbar}{2}\qty[\qty(\hat{a}_{\omega}^\dagger)^2+\hat{a}_{\omega}^2+1]\ ,
\end{equation}
where we made use of $\partial_{\omega} a^\dagger_\omega = \hat{a}_{\omega} / 2\omega$ and $\partial_{\omega} \hat{a}_{\omega} = \hat{a}_{\omega}^\dagger / 2\omega$, which can be seen from their representation in the Fock state basis $\ket{n}_\omega$ . With the help of 
\begin{equation}
	\expp{-i\psi \hat{a}_\omega^\dagger \hat{a}_\omega} \hat{a}_\omega \expp{i\psi \hat{a}_\omega^\dagger \hat{a}_\omega} 
		= \expp{i \psi} \hat{a}_\omega
\end{equation}
we get
\begin{eqnarray}
	&V&(\alpha) \pdv{\mathcal{H}_\omega}{\omega}V^\dagger(\alpha)\nonumber\\
		&=& \hbar\hat{a}^\dagger_\omega \hat{a}_{\omega} + \frac{\hbar}{2}\qty[\expp{-2 i \alpha \omega t}\qty(\hat{a}_{\omega}^\dagger)^2+\expp{2 i \alpha \omega t}\hat{a}_{\omega}^2+1]\ .
\end{eqnarray}
Insertion and subsequent integration provides the local generator
\begin{equation}
	\mathscr{K} 
		= t \qty(\hat{a}^\dagger_\omega \hat{a}_{\omega} + \frac{1}{2}) - \frac{i}{4  \omega} \qty[\qty(1-\expp{-2i \omega t})\hat{a}_{\omega}^2+\qty(1-\expp{2i \omega t})\hat{a}_{\omega}^{\dagger 2}].
\end{equation}
Next, the QFI is calculated. 
The annihilation and creation operator $\hat{c}_{\omega}$ and $\hat{c}^\dagger_{\omega}$, defined by
\begin{eqnarray}
	\hat{c}_\omega 
		&=& S_{\omega}^\dagger\qty(r\expp{i \chi}) D_{\omega}^\dagger\qty(\alpha) R_{\omega}^\dagger\qty(\psi) \hat{a}_{\omega} R_{\omega}\qty(\psi) D_{\omega}\qty(\alpha) S_{\omega}\qty(r\expp{i \chi})\nonumber\\
		&=& \expp{i \psi}\qty(\cosh(r) \hat{a}_{\omega}+ \expp{i \chi}\sinh(r) \hat{a}_{\omega}^\dagger+\alpha)
\end{eqnarray}
and $\qty(\hat{c}_{\omega})^\dagger=\hat{c}^\dagger_{\omega}$, can be used to rewrite the expectation values of $\mathscr{K}$ as follows
\begin{equation}
	\ev{\mathscr{K}^k}{\psi_0}
		=\ev{\eval{\mathscr{K}^k}_{\substack{\hat{a}^\dagger=\hat{c}^\dagger\\\hat{a}=\hat{c}}}}{0}.
\end{equation}
Using the formulae $\hat{a}\ket{n}=\sqrt{n}\ket{n-1}$ and $\hat{a}^\dagger\ket{n}=\sqrt{n+1}\ket{n+1}$, 
we obtain the QFI after a straightforward calculation:
\begin{widetext}
\begin{eqnarray}\label{eq:QFI-pure-state-final}
	\qfi{\rho(t)}{\omega}
		&=& \frac{2t}{\omega}\sin\omega t \bigg[4\alpha^2\cos(2\psi-\omega t)\cosh 2r+ \cos(\chi+2\psi-\omega t)\qty(4\alpha^2 \sinh 2r + \sinh 4r)\bigg]\nonumber\\
		& &+ \frac{2}{\omega^2}\sin^2\omega t \bigg[\sinh^2 2r\cos[2](\chi+2\psi-\omega t) + 1+ 2\alpha^2\qty(\cosh 2r + \cos(\chi+4\psi-2\omega t)\sinh 2r)\bigg]\nonumber\\
		& &+ 2 t^2 \qty[2\alpha^2\qty(\cosh 2r+\cos \chi\sinh 2r)+\sinh^2 2r].
\end{eqnarray}
\end{widetext}
Comparison with \eq \eqref{eq:QFI-undamped} for a pure Gaussian state, \ie $\Nth=0$, shows that the results are identical. 
\section{Calculation of QFI\label{App:QFI-Calculation}}
Here we report the calculation of the QFI for measuring $\omega$. First, the dynamics resulting from ME \eqref{eq:ME} is determined. Then the QFI for the undamped case is calculated. Finally, the exact QFI for the damped case is given.

The solutions of ME \eqref{eq:ME} are given by {\cite{Isar2006}}
\begin{subequations}\label{eq:ME-Solution-1-ord}
\begin{eqnarray}
	\ev{q}_t 
		&=& \expp{-\frac{\gamma t}{2}}\qty[\cos(\omega t)\ev{q}_0+\frac{1}{M\omega} \sin(\omega t)\ev{p}_0], \\
	\ev{p}_t 
		&=& \expp{-\gamma t/2}\qty\bigg[\cos(\omega t)\ev{p}_0-M\omega\sin(\omega t)\ev{q}_0]\quad\ \label{eq:p(t)}
\end{eqnarray}
\end{subequations}
and 
\begin{widetext}
\begin{subequations}\label{eq:ME-Solution-2-ord}
\begin{eqnarray}
	\sigma_{qq}(t) 
		&=& \frac{\hbar}{2M\omega}\qty(1+2\nbar)\qty(1-\expp{-\gamma t})\nonumber\\
		& &+\expp{-\gamma t} \qty[\cos[2](\omega t)\sigma_{qq}(0)+\frac{\sin[2](\omega t)}{M^2\omega^2}\sigma_{pp}(0) +\frac{\sin(2 \omega t)}{M\omega} \sigma_{pq}(0)],\\
	\sigma_{pp}(t) 
		&=& \frac{\hbar M \omega}{2}\qty(1+2\nbar)\qty(1-\expp{-\gamma t})\nonumber\\
		& &+\expp{-\gamma t} \qty\bigg[\cos[2](\omega t)\sigma_{pp}(0)+M^2\omega^2\sin[2](\omega t)\sigma_{qq}(0) -M\omega\sin(2 \omega t) \sigma_{pq}(0)],\\
	\sigma_{pq}(t) 
		&=& \expp{-\gamma t} \qty[\cos(2\omega t)\sigma_{pq}(0)+\frac{1}{M \omega}\sin(\omega t)\cos(\omega t) \qty\Big(\sigma_{pp}(0)-M^2\omega^2\sigma_{qq}(0))].
	\end{eqnarray}
\end{subequations}
\end{widetext}
{Indeed, the second equation \eqref{eq:p(t)} is an immediate consequence of $p = M\partial_t q$.}
For the general single-mode Gaussian state in \eq \eqref{eq:Gaussian-State-Representation} the initial expectation values are given by
\begin{subequations}\label{eq:Initial-Exp-1-ord}
	\begin{eqnarray}
	\ev{q}_0 
		&=& \alpha \sqrt{\frac{2 \hbar}{M \omega_0}} \cos(\psi), \\
	\ev{p}_0 
		&=& \alpha \sqrt{2 \hbar M \omega_0}\sin(\psi). \\
	\sigma_{qq}(0)
		&=&\frac{\hbar}{2 M \omega_0}(2\Nth+1)\nonumber\\
		& &\qty[\cosh(2r)+\cos(\chi+2\psi)\sinh(2r)],\\
	\sigma_{pp}(0)
		&=&\frac{\hbar M \omega_0}{2 }(2\Nth+1)\nonumber\\
		& &\qty[\cosh(2r)-\cos(\chi+2\psi)\sinh(2r)],\\
	\sigma_{pq}(0)
		&=& \frac{\hbar}{2}(2\Nth+1)\sin(\chi+2\psi)\sinh(2r).	
	\end{eqnarray}
\end{subequations}
Here we give the expectation values with respect to the initial frequency $\omega_0$. The time evolution of the ME \eqref{eq:ME}, on the other hand, is with respect to the new frequency $\omega$, as described in the scheme.
\subsection{Undamped case}
We start with the calculation for the QFI of the undamped case of \eq \eqref{eq:QFI-undamped}. The undamped dynamic corresponds to the expectation values from \eq \eqref{eq:ME-Solution-1-ord} and \eq \eqref{eq:ME-Solution-2-ord} for $\gamma\to0$. By using these results we calculated the five parameters of interest $\Sigma^{-1},\partial_{\omega}\Sigma, P, \partial_{\omega}P, \partial_{\omega}\ev{\Xvec}$. For the sake of clarity we give the results after executing the limit $\omega_0\to\omega$ and additionally use the dimensionless time $\tau=\omega t$. The derivative of quadrature operator is given by
\begin{equation}\label{eq:App-XDerivative}
	\partial_{\omega}\ev{\Xvec} 
		=\alpha\sqrt{\frac{2  M\hbar}{\omega}}\begin{pmatrix}
		 \frac{\tau\sin(\psi-\tau)-\sin(\psi)\sin(\tau)}{M \omega}\\
		- \tau\cos(\psi-\tau)-\cos(\psi)\sin(\tau)
		\end{pmatrix} 
\end{equation}
The purity and its derivative read
\begin{subequations}
\begin{eqnarray}
	P
		&=& \frac{1}{1+2\Nth},\\
	\partial_{\omega} P
		&=& \frac{2 \Nth (1+\Nth) \ln(1+1/\Nth)}{\omega(1+2\Nth)^2},
\end{eqnarray}
\end{subequations}
whereas the derivative of the covariance matrix is described by the following equations:
\begin{widetext}
\begin{subequations}\label{eq:App-sigmaDerivative}
	\begin{eqnarray}
	\frac{2 M \omega}{\hbar}\partial_{\omega}\sigma_{qq}(t)
		&=& \frac{1}{\omega P}\qty[-2\cosh(2r)\sin[2](\tau)+\qty(\cos(\chi+2\psi)-\cos(2\tau-\chi-2\psi)-2\tau\sin(2\tau-\chi-2\psi))\sinh(2r)]\nonumber\\
		& &-\frac{\partial_{\omega} P}{P^2}\qty[\cosh(2r)+\cos(2\tau-\chi-2\psi)\sinh(2r)],\\
	\frac{2}{\hbar M  \omega}\partial_{\omega}\sigma_{pp}(t)
		&=& \frac{1}{\omega P}\qty[2\cosh(2r)\sin[2](\tau)+\qty(\cos(\chi+2\psi)-\cos(2\tau-\chi-2\psi)+2\tau\sin(2\tau-\chi-2\psi))\sinh(2r)]\nonumber\\
		& &-\frac{\partial_{\omega} P}{P^2}\qty[\cosh(2r)-\cos(2\tau-\chi-2\psi)\sinh(2r)],\\
	\frac{2}{\hbar} \partial_{\omega}\sigma_{pq}(t)
		&=& \frac{\partial_{\omega} P}{P^2}\sin(2\tau-\chi-2\psi)\sinh(2r) - \frac{1}{ \omega P} \qty(\sin(2\tau)\cosh(2r)+2\tau\cos(2\tau-\chi-2\psi)\sinh(2r)).
	\end{eqnarray}
\end{subequations}
\end{widetext}
By inserting \eq \eqref{eq:App-XDerivative}-\eqref{eq:App-sigmaDerivative} into \eq \eqref{eq:General-QFI} one obtain \eq \eqref{eq:QFI-undamped}.
\subsection{Damped case}
By repeating the previous calculations with a non-vanishing $\gamma$, we estimate the QFI for damped Gaussian states for measuring $\omega$. Since the solution is too heavy, we specify the three terms of \eq \eqref{eq:General-QFI} separately{, \ie} 
\begin{equation}\label{eq:App-QFI-3parts}
\qfi{\rho(t)}{\omega} = \QFI_{1,\omega}+ \QFI_{2,\omega} + \QFI_{3,\omega}, 
\end{equation}
where 
\begin{subequations}
\begin{eqnarray}
	\QFI_{1,\omega}
		&=&(2(1+P^2))^{-1}\tr\qty[\qty(\Sigmavec^{-1} \partial_\theta\Sigmavec)^2], \\
	\QFI_{2,\omega} 
		&=&2  \qty(\partial_\theta P)^2 (1- P^4)^{-1}, \\
	\QFI_{3,\omega} 
		&=& \qty(\partial_\theta\ev{\Xvec})^\trans \Sigmavec^{-1}\partial_\theta\ev{\Xvec}.
\end{eqnarray}
\end{subequations}
Repeating the previous calculation for a non-vanishing damping leads to the following exact result of the QFI:
\begin{widetext}
\begin{eqnarray}
	\QFI_{1,\omega}
		&=&\frac{P^4(\tau)}{2\omega^2\expp{4g\tau}\qty(1+P^2(\tau))}\Bigg\{
		8 A_1^2\qty(A_1^2 + a_{1,\tau}^2 +2 a_{1,\tau} A_1 \ch)\sh^2\tau^2\nonumber\\
		& &+8 A_1\qty(A_1^2 + a_{1,\tau}^2 +2 a_{1,\tau}  A_1 \ch)\qty[A_1 \sin(\xi) \ch + (a_{1,\tau}+A_1 \ch)\sin(2\tau-\xi)]\sh\tau\nonumber\\
		& &+ \frac{A_1^2}{2} \Big( 4  A_1^2 \left(\sh^2+2\right)
		+ A_2^2  A_3^2+2  A_2 A_3 a_{2,\tau}a_3   \ch+a_{2,\tau}^2 a_3^2 \left(2 \sh^2+1\right)\nonumber\\
			& &\phantom{+ \frac{A_1^2}{2} \Big(}-2 A_1 \qty[ 4 A_1 \cos(2\tau) + A_1(\cos(2\xi)+\cos(4\tau-2\xi)+4\sin[\xi]\sin[2\tau-\xi])\sh^2-4a_{2,\tau} a_3 \sin(\tau)\sin(\tau-\xi)\sh] \nonumber\\
		& &+  A_1 a_{1,\tau} \Big(2 A_1 \left[ 4 A_1 \ch \sin[2](\tau)\left(3+2\cos[2](\tau-\xi)\sh^2\right)+\sh\cos(2\tau-\xi)\qty( A_2  A_3-2  a_{2,\tau}a_3 \ch)-2  A_2  A_3 \sh \cos(\xi) \right]\nonumber\\
			& &\phantom{+ 4  A_1 a_1(\tau) \Big(}+ A_2^2  A_3^2 \ch+2  A_2    A_3 a_{2,\tau}a_3+ a_{2,\tau}^2 a_3^2 \ch\Big)\nonumber\\
		& &+\frac{a_{1,\tau}^2}{2} \Big( 4 A_1^2 (7+6\sh^2)+a_{2,\tau}^2 a_3^2+A_2 A_3 \qty[2 a_{2,\tau}a_3 \ch + A_2 A_3 (1+2\sh^2)] \nonumber\\
		& &\phantom{+\frac{a_{1,\tau}^2}{2} \Big(} + 2 A_1 \big\{ A_1 \sh^2 \qty[2 \qty(\cos(2\tau-2\xi) - 5 \cos(2\tau)) - \cos(4\tau-2\xi) + \cos(2\xi)]- 12A_1 \cos(2\tau) \nonumber\\
		& & \phantom{+\frac{a_{1,\tau}^2}{2} \Big( + 2 A_1 \big\{}	+ 8 A_2 A_3 \ch \sh \sin(\tau) \sin(\tau-\xi) - 4 a_{2,\tau} a_3 \sh \cos(\tau) \cos(\tau-\xi)	\big\}\Big)\nonumber\\
		& &+a_{1,\tau}^3 \qty[2  A_2  A_3 \sh \cos(2\tau-\xi)-4  A_1 \ch (\cos(2\tau)-2)]+2 a_{1,\tau}^4 \Bigg\}, \\
	\QFI_{2,\omega}
		&=& \frac{\expp{-4g\tau} P^6(\tau)}{2\omega^2\qty(1-P^4(\tau))}\big[A_1 A_2 A_3 + a_{1,\tau}a_{2,\tau} a_3 + \qty(A_1 a_{2,\tau} a_3 + a_{1,\tau} A_2 A_3)\ch-2 a_{1,\tau} A_1 \cos(\xi)\sh\big]^2,\\
	\QFI_{3,\omega} 
		&=& \frac{4 \expp{-2g\tau}\alpha^2P^2(\tau)}{\omega^2}\big\{\qty[a_{1,\tau}+A_1 \qty(\ch+\cos(\chi)\sh)]\tau^2+\qty[\cos(\tau-2\psi)\qty(a_{1,\tau}+A_1 \ch)+A_1 \cos(\tau-\xi)\sh]2\tau\sin(\tau)\nonumber\\
		& &+\qty[a_{1,\tau}+A_1 \qty(\ch+\cos(2\tau-\xi-2\psi)\sh)]\sin[2](\tau)\big\},
\end{eqnarray}
\end{widetext}
where we introduced a new angle $\xi=\chi+2\psi$, $\ch =\cosh(2r)$, $\sh = \sinh(2r)$ and 
\begin{subequations}
	\label{eq:App-QFI-abbr}
	\begin{align}
	a_1 		&= 1+2\nbar, \ 			&A_1		&= 1+2\Nth,\\
	a_2 		&= 4\nbar(1+\nbar),\ 	& A_2 		&= 4\Nth(1+\Nth),\\	
	a_3 		&= \ln(1+1/\nbar), \ 	&A_3 		&=\ln(1+1/\Nth), \\
	a_{1,\tau} 	&=(\expp{g \tau}-1)a_1,\ &a_{2,\tau}&= (\expp{g \tau}-1)a_2.
	\end{align}
\end{subequations}

{Finally the maximum of the QFI of an initial ground state, which was used to approximate the QFI of the coherent state, is determined. For an initial ground state, the QFI simplifies to}
\begin{widetext}
\begin{equation}
	\omega^2\qfi{\rho}{\omega} = 
		\frac{1+\qty[\expp{g \tau}\qty(1+2\nbar)-2\nbar]^2 - 2 \qty[\expp{g\tau}\qty(1+2\nbar)-2\nbar]\cos(2\tau)}{2\qty[2\nbar^2 - 2\expp{g\tau}\nbar (1+2\nbar)+\expp{2g\tau}\qty(1+2\nbar+2\nbar^2)]}
		+ \frac{\qty(\expp{g\tau}-1)\nbar\qty(1+\nbar)^2 \ln[2](1+1/\nbar)}{\expp{g\tau}(1+\nbar)-\nbar}.
\end{equation}
\end{widetext}
{Numerical maximization of $\qfi{\rho}{\omega}$ with respect
  to the three parameters $\tau, g,\nbar$ returns the value $2.135/\omega^2$.}
\bibliography{apsbib}
\end{document}